# ON THE WIEDEMANN-FRANZ LAW IN THERMOELECTRIC COMPOSITES

*A.A. Snarskii[1], M. I. Zhenirovskii[2], I. V. Bezsudnov[3]*

*[1] – National Technical University of Ukraine «KPI», Kiev, Ukraine;*
*[2] – Nikolay Bogoliubov Institute of Theoretical Physics, Kiev, Ukraine;*
*[3] – Nauka Service Ltd, Moscow, Russia*

*Theoretical analysis of the effect of thermoelectric phenomena on the relation of effective electrical and thermal conductivity in the macroscopically inhomogeneous media is carried out. Plane-layered structures, two-dimensional self-dual and three-dimensional randomly inhomogeneous media are examined. It is shown that the relation of effective electrical to thermal conductivity, which is independent of concentration in so-called Wiedemann –Franz media, in the case of high values of internal thermoelectric figure of merit can be considerably different from the one for the local values in each phase. The decrease of the relation may have negative impact on the effective figure of merit of composites and thus lead to the decrease of thermoelectric devices efficiency.*

## Introduction

The effective kinetic coefficients connecting average by volume thermodynamic forces and flows, are fundamental parameters of macroscopically inhomogeneous composites. For thermoelectric phenomena these relationships take the form

$$\begin{aligned} \langle \mathbf{j} \rangle &= \sigma_e \langle \mathbf{E} \rangle - \sigma_e \alpha_e \langle \nabla T \rangle, \\ \langle \mathbf{q} \rangle &= -\kappa_e \left( 1 + \frac{\sigma_e \alpha_e^2}{\kappa_e} T \right) \langle \nabla T \rangle + \sigma_e \alpha_e T \langle \mathbf{E} \rangle, \end{aligned} \tag{1}$$

where $\sigma_e$, $\kappa_e$ are effective electrical and thermal conductivity, respectively, $\alpha_e$ is thermoEMF, $\langle \mathbf{E} \rangle$ and – $\langle \nabla T \rangle$ are average by volume electric field intensity and temperature gradient, respectively, while $\langle \mathbf{j} \rangle$ and $\langle \mathbf{q} \rangle$ respectively denote average by volume electric current density and heat flow density.

Three kinetic coefficients $\sigma$, $\kappa$ and $\alpha$ allow to introduce dimensionless temperature in a natural way

$$\tilde{T} = \frac{\sigma \alpha^2}{\kappa} T = ZT . \tag{2}$$

This dimensionless temperature and, correspondingly, expression $Z = \sigma\alpha^2/\kappa$ [1], known as thermoelectric figure of merit, acts as determining factor when describing energy processes in thermoelectric media – efficiency, refrigerating capacity, etc.

The Wiedemann-Franz law, which describes stability of the relation $\sigma/\kappa$ at $T$=const for different metals, substantially limited the search for metallic conductors with high figure of merit. Following (2), increase of $Z$ becomes possible due to increase of electrical conductivity and simultaneous decrease of thermal conductivity. This suggestion does not conform to the Wiedemann-Franz law. Specifically, the nonfulfillment of the Wiedemann-Franz law in semiconductors to a considerable extent allowed to create thermoelectric devices possessing high figure of merit related to high efficiency [2]. When in search for new materials (e.g. obtained by alloying of semiconductors), besides the increase in thermoEMF required changes in the relation $\sigma/\kappa$ are also obtained.

In thermoelectric composites effective figure of merit is the basic energy parameter

$$Z_e = \frac{\sigma_e \alpha_e^2}{\kappa_e}, \tag{3}$$

which brings forward natural and most important question about the dependence of $\sigma_e / \kappa_e$ on the local values of effective kinetic coefficients (EKC) of independent phases forming the composite ($\sigma_1, \kappa_1, \alpha_1, \sigma_2, ...$), concentrations of these phases, and geometric structure of composite phase arrangement. In particular, we can formulate the following question: will the composite comply with the Wiedemann-Franz law, i.e. will it satisfy the relation $\sigma_e / \kappa_e = \text{const}$, if phases are considered to be Wiedemann-Franz, specifically $\sigma_1 / \kappa_1 = \sigma_2 / \kappa_2 = ...$ Positive answer is indubitable in the absence of thermoelectric phenomena. This work is devoted to finding the answer to this question for the case when thermoelectric phenomena are significant and some consequences of this answer which are important for thermoelectric materials science.

Several types of two-phase thermoelectric composites with different geometric structure, i.e. plane-layered and self-dual media for which it is possible to obtain precise analytical expressions of effective kinetic coefficients will be examined below. Randomly inhomogeneous three-dimensional media which can be analyzed only within the framework of specific approximations will also be examined. The approximation of so-called self-congruent field (average field, the Bruggemann-Landauer method) and the approach of theory of flow, possible with strong inhomogeneity of phases in local conductivity and with concentration, close to the threshold of flow is selected.

Let us immediately note that the effect of thermoelectric phenomena on relation $\sigma_e / \kappa_e$ occurs only in the medium with inhomogeneous thermoelectric properties, i.e. when $\alpha_1 \neq \alpha_2$. In the composite which is inhomogeneous in electrical conductivity ($\sigma_1 \neq \sigma_2$) and (or) in thermal conductivity ($\kappa_1 \neq \kappa_2$), but homogeneous in thermoelectric properties ($\alpha_1 = \alpha_2$), relation $\sigma_e / \kappa_e$ will be the same as in the absence of thermoelectric phenomena

The current state of thermoelectric materials science, in particular the values of material constants of thermoelectric materials with a comparatively high figure of merit is reflected in the reference book [2].

The aim of the present paper is to analyze the effect of thermoelectric phenomena on relation $\sigma_e / \kappa_e$, which to a considerable extent determines figure of merit of macroscopically inhomogeneous materials.

## 1. Plane-layered media

Effective kinetic coefficients for two-phase media in the plane-layered media were obtained in [3] (see also [4]), general expressions for the arbitrary one-dimensional coordinate dependence they are given in [5]:

$$\sigma_{xx}^e(p) = \frac{\sigma_1 \sigma_2}{p\sigma_2 + (1-p)\sigma_1} \frac{1}{1+\tilde{Z}(p)T}, \quad \sigma_{zz}^e(p) = p\sigma_1 + (1-p)\sigma_2, \tag{4}$$

$$\kappa_{xx}^e(p) = \frac{\kappa_1 \kappa_2}{p\kappa_2 + (1-p)\kappa_1}, \quad \kappa_{zz}^e(p) = \left[p\kappa_1 + (1-p)\kappa_2\right]\left(1+\tilde{Z}(1-p)T\right), \tag{5}$$

$$\alpha_{xx}^e(p) = \frac{\kappa_2 \alpha_1 p + \kappa_1 \alpha_2 (1-p)}{p\kappa_2 + (1-p)\kappa_1}, \quad \alpha_{zz}^e(p) = \frac{\alpha_1 \sigma_1 p + \alpha_2 \sigma_2 (1-p)}{p\sigma_1 + (1-p)\sigma_2}. \tag{6}$$

Here $x$ axis takes perpendicular direction to the layers, $\tilde{Z}(p)$ is so called intrinsic figure of merit [6] which in this case of plane-layered media is equal to

$$\tilde{Z}(p)T = p(1-p)\frac{\sigma_1\sigma_2(\alpha_1-\alpha_2)^2}{\left[p\sigma_2+(1-p)\sigma_2\right]\left[p\kappa_1+(1-p)\kappa_2\right]}T. \tag{7}$$

As can be seen from (4) and (5), the value of intrinsic figure of merit $\tilde{Z}(p)$ can introduce significant changes into the components of electrical and thermal conductivity tensors $\sigma^e_{xx}$ and $\kappa^e_{xx}$, respectively. In the case when the first phase is metal (high electrical conductivity), and the second phase is semiconductor (low in comparison with the metal conductivity) $\tilde{Z}(p)T$ has a sharp peak near $p=1$. The value of this peak can be estimated as follows: for the case of Wiedemann -Franz composite, i.e. when intrinsic figure of merit can be written in the form

$$\tilde{Z}_{WF}(p)T = \frac{\sigma_1}{\kappa_1}(\alpha_1-\alpha_2)^2 T\frac{p(1-p)}{\left[p\sigma_1/\sigma_2+(1-p)\right]^2}, \tag{8}$$

where subscript *WF* accompanying the symbol of intrinsic figure of merit means that the medium in question is Wiedemann –Franz medium.

Maximum value of $\tilde{Z}_{WF}(p)T$ is achieved when $p=h/(1+h)$, where $h=\sigma_2/\sigma_1$ and is equal to

$$\max\tilde{Z}_{WF}(p)T = \tilde{Z}_{WF}\left(p=h/(1+h)\right)T = \frac{1}{4}\frac{\sigma_1}{\kappa_1}(\alpha_1-\alpha_2)^2 T\frac{\sigma_1}{\sigma_2}, \tag{9}$$

which means the more pronounced is inhomogeneity in electrical conductivity ($\sigma_1/\sigma_2$), the higher is the value of intrinsic figure of merit $\tilde{Z}_{WF}$.

With the increase in inhomogeneity the value of $\tilde{Z}_{WF}$ unlimitedly grows and can considerably exceed unity, which means that $\sigma^e_{xx}$ (4) and $\kappa^e_{zz}$ (5) can substantially change due to the thermoelectric phenomena. Let us note that intrinsic figure of merit exhibits opposite behaviour in $\sigma^e_{xx}$ and $\kappa^e_{zz}$: $\sigma^e_{xx}$ decreases with the increase in $\tilde{Z}_{WF}$, while $\kappa^e_{zz}$ increases; therefore deviation from the Wiedemann-Franz law increases with the increase of intrinsic figure of merit. Fig. 1 shows the dependences of electrical to thermal conductivity ratio on the concentration and value of thermoEMF of the second phase.

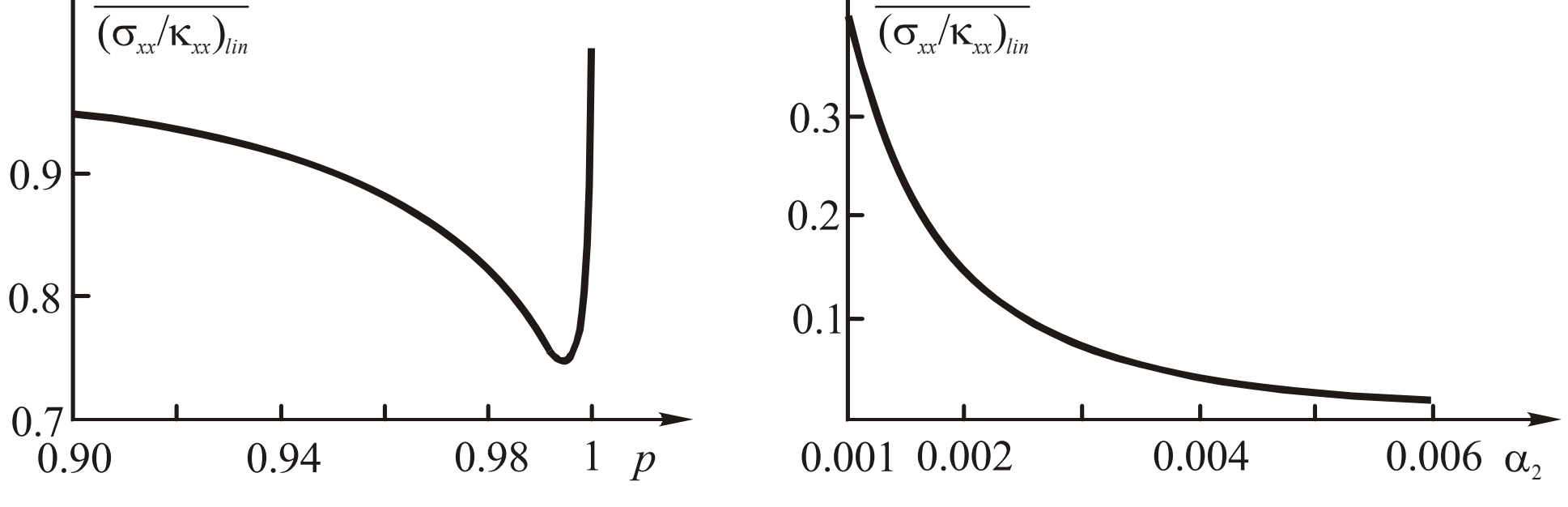

*Fig. 1. Plane-layered media. The first phase is constituted by metal, and the second phase is constituted by semiconductor.*
*By way of example, the following parameter values are selected: $\alpha_1=0$, $\kappa_1=1.2\cdot10^2$, $\sigma_1=6.029\cdot10^6$, $\kappa_2=6.37\cdot10^{-1}$, $\sigma_2=3.2\cdot10^4$. ThermoEMF of the second phase varies within $2\cdot10^{-4} \div 6\cdot10^{-3}$. For descriptive reasons, the value of ratio ($\sigma_{xx}/\kappa_{xx}$) is rated for this ratio in the absence of thermoelectric phenomena – $(\sigma_{xx}/\kappa_{xx})_{lin}$.*

For a plane-layered composite, which has the first phase with poor conductivity (dielectric), the parameter characteristic of inhomogeneity is $h >> 1$, while intrinsic figure of merit decreases with the increase in $h$. Thus, deviation from the Wiedemann-Franz law becomes insignificant. In terms of quality, this phenomenon can be attributed to the fact that a change in the effective electrical conductivity and thermal conductivity is associated with heat flows generated by eddy thermoelectric currents. In the case of "metal-semiconductor" composites emerging thermoelectric fields generate eddy thermoelectric currents which are closed through the metal phase. In the case of "semiconductor-dielectric" composites poorly conducting phase prevents closing of currents within the composite.

## 2. Self-dual media

There exists a class of two-dimensional media for which it is possible to obtain most precise solution of effective kinetic coefficients problem applicable to whatever large inhomogeneity [7]. This class also includes two-dimensional randomly inhomogeneous media on the threshold of flow. Effective kinetic coefficients of self-dual media in the case of thermoelectric phenomena were obtained in [8] and [9], and here we will present them in the form convenient for future reference

$$\sigma_e = \frac{\sigma_D^e}{\sqrt{1+\tilde{Z}T}}, \quad \kappa_e = \kappa_D^e\sqrt{1+\tilde{Z}T}, \quad \alpha_e = \frac{\alpha_1\sqrt{\sigma_1\kappa_2}+\alpha_2\sqrt{\sigma_2\kappa_1}}{\sqrt{\sigma_1\kappa_2}+\sqrt{\sigma_2\kappa_1}}, \tag{10}$$

where

$$\sigma_D^e = \sqrt{\sigma_1\sigma_2}, \quad \kappa_D^e = \sqrt{\kappa_1\kappa_2}, \tag{11}$$

are effective electrical and thermal conductivities in the absence of thermoelectric phenomena, first obtained in [7], and the value

$$\tilde{Z}T = \frac{\sigma_D^e}{\kappa_D^e}\frac{(\alpha_1-\alpha_2)^2}{\left(\sqrt{\frac{\sigma_2}{\sigma_1}}\Big/\sqrt{\frac{\kappa_2}{\kappa_1}}+\sqrt{\frac{\kappa_2}{\kappa_1}}\Big/\sqrt{\frac{\sigma_2}{\sigma_1}}\right)}\cdot T \tag{12}$$

represents intrinsic figure of merit of self-dual media [6].

As it can be seen from (10), relation $\sigma_e/\kappa_e$ is greatly dependent on intrinsic figure of merit $\tilde{Z}T$ and therefore on thermoelectric phenomena

$$\frac{\sigma_e}{\kappa_e} = \frac{\sigma_D^e}{\kappa_D^e}\frac{1}{1+\tilde{Z}T}. \tag{13}$$

Especially simple form is taken by (13) in the case of Wiedemann-Franz composite when $\sigma_1/\kappa_1 = \sigma_2/\kappa_2$. In this case

$$\tilde{Z}_{WF}T = \frac{1}{4}\frac{\sigma_D^e}{\kappa_D^e}(\alpha_1-\alpha_2)^2 T, \qquad \left.\frac{\sigma_D^e}{\kappa_D^e}\right|_{WF} = \frac{\sigma}{\kappa}, \tag{14}$$

where $\sigma/\kappa = \sigma_1/\kappa_1 = \sigma_2/\kappa_2$ and (13) takes the form

$$\frac{\sigma_e}{\kappa_e} = \frac{\sigma}{\kappa}\frac{1}{1+\frac{1}{4}\frac{\sigma}{\kappa}(\alpha_1-\alpha_2)^2 T}, \tag{15}$$

whence it immediately follows that

$$\frac{\sigma_e}{\kappa_e} \le \frac{\sigma}{\kappa}, \tag{16}$$

and that the ratio $\sigma_e/\kappa_e$ is inversely proportional to the composite inhomogeneity in thermoEMF $(\alpha_1-\alpha_2)^2$.

## 3. Three-dimensional randomly inhomogeneous media

For the randomly inhomogeneous media it is possible to obtain the solution of EKC problem in the entire range of concentration variations only in the approximate form. In [10] so-called method of isomorphism was proposed. In certain cases this method allows to reduce the problem of effective kinetic coefficients of thermoelectric composite to the simpler problem of effective electrical conductivity in the absence of thermoelectric phenomena. In [11] this method is developed and applied, in particular, to thermoelectric composites with consideration to the effect of thermoelectric phenomena on the electrical and thermal conductivity. According to [11]

$$\left.\begin{aligned}
\sigma_e &= \frac{(\mu\sigma_1-\sigma_2)f_\lambda-(\lambda\sigma_1-\sigma_2)f_\mu}{\mu-\lambda}\\
\alpha_e &= \frac{(\mu\sigma_1\alpha_1-\sigma_2\alpha_2)f_\lambda-(\lambda\sigma_1\alpha_1-\sigma_2\alpha_2)f_\mu}{(\mu\sigma_1-\sigma_2)f_\lambda-(\lambda\sigma_1-\sigma_2)f_\mu}\\
\kappa_e &= \frac{\sigma_1\kappa_1(\mu-\lambda)f_\lambda f_\mu}{(\mu\sigma_1-\sigma_2)f_\lambda-(\lambda\sigma_1-\sigma_2)f_\mu}
\end{aligned}\right\}, \tag{17}$$

where

$$\begin{Bmatrix}\mu\\ \lambda\end{Bmatrix} = \frac{1}{4\sigma_1\kappa_1}\left\{\left[(\sqrt{\sigma_1\kappa_2}+\sqrt{\sigma_2\kappa_1})^2+\sigma_1\sigma_2T(\alpha_1-\alpha_2)^2\right]^{\frac{1}{2}} \pm \left[(\sqrt{\sigma_1\kappa_2}-\sqrt{\sigma_2\kappa_1})^2+\sigma_1\sigma_2T(\alpha_1-\alpha_2)^2\right]^{\frac{1}{2}}\right\}^2. \tag{18}$$

Functions $f_\lambda$ and $f_\mu$ are determined as follows. Let us write down effective electrical conductivity $\sigma_e$ (without taking into account thermoelectric phenomena) of the considered system in the form $\sigma_e=\sigma_1 f(p,h)$, where $h=\sigma_2/\sigma_1$, $\sigma_1$ and $\sigma_2$ are electrical conductivities of the first and the second components, respectively, $p$ is concentration of the first component. Then functions $f_\lambda$ and $f_\mu$ are obtained from $f(p,h)$ by substitution $h\to\lambda$ и $h\to\mu$.

Therefore

$$f_\lambda = f(p,\lambda), \qquad f_\mu = f(p,\mu). \tag{19}$$

Thus, for three-dimensional randomly inhomogeneous medium mean fields approximation (EMA-approximation) [12, 13] yields (see Figs. 2 and 3):

$$f=\frac{\sigma_e}{\sigma_1}=\frac{1}{4}\left\{\left[3p(1-h)+2h-1\right]+\sqrt{\left[3p(1-h)+2h-1\right]^2+8h}\right\}. \tag{20}$$

Expression (17) is true both for two-dimensional and three-dimensional isotropic two-component systems of arbitrary structure. All information about the form of inclusions and their arrangement is included in the function $f$. If for any two-component conducting system free of thermoelectric phenomena $f(p,h)$ is known for all $p$ and $h$ values, then formula (17) provides complete solution of the problem of thermoelectric properties of this system.

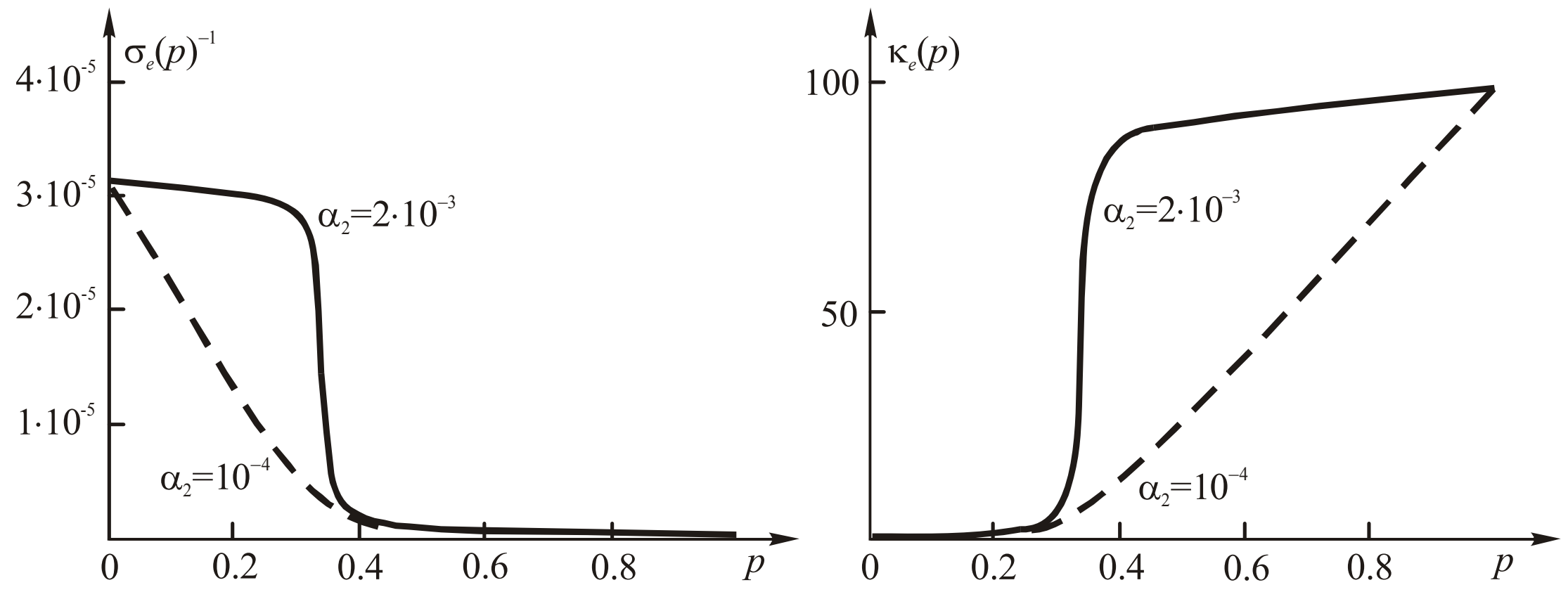

*Fig. 2. Randomly inhomogeneous media. EMA-approximation. The first phase is constituted by metal; the second phase is constituted by semiconductor. By way of example, the following parameter values are selected: $\kappa_1$=99.314, $\sigma_1$=5·10$^6$, $\kappa_2$=0.637, $\sigma_2$=3.207·10$^4$. ThermoEMF of the second phase varies within 10$^{-4}$ ÷ 20·10$^{-4}$.*

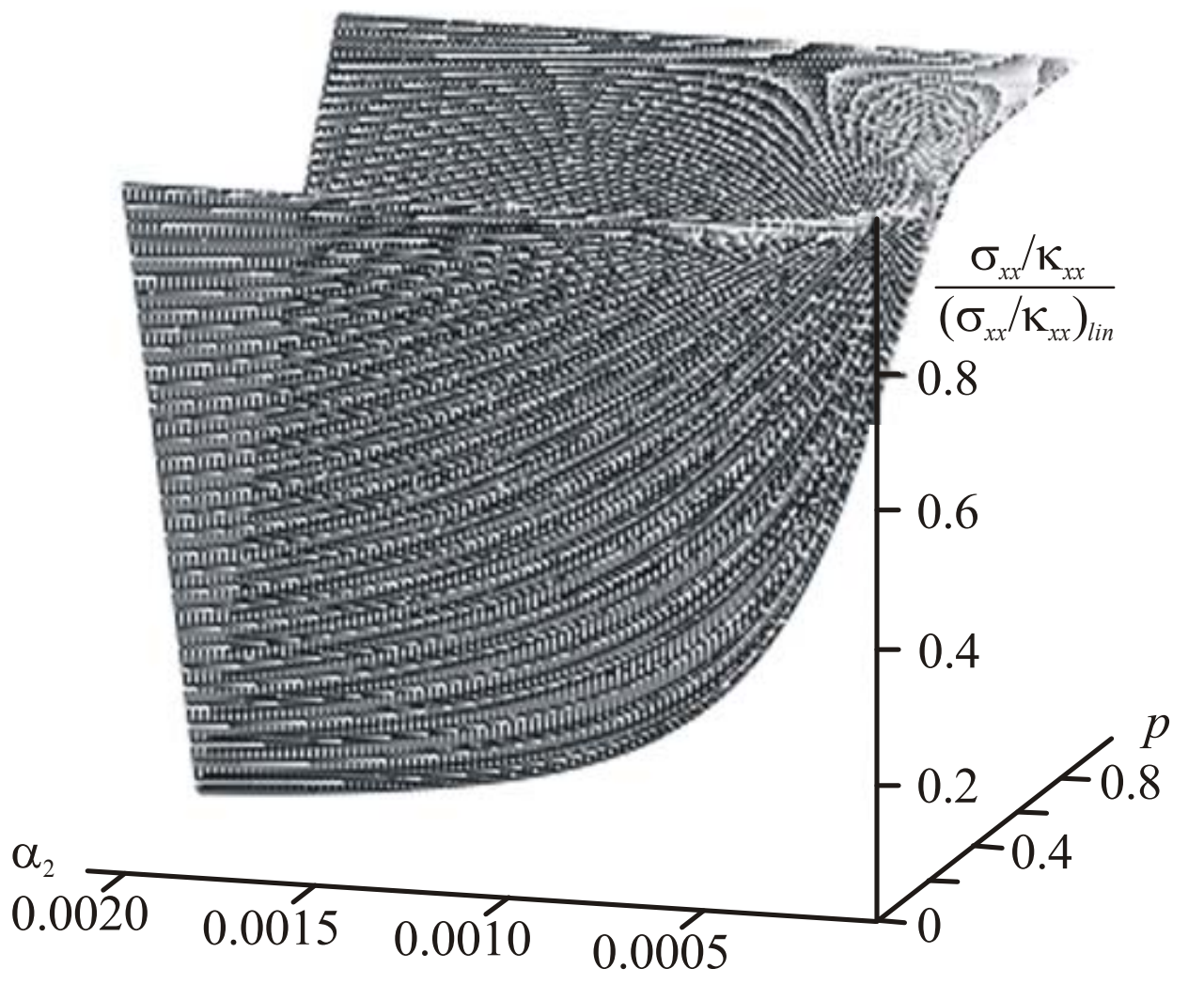

*Fig. 3. Randomly inhomogeneous media. EMA-approximation. Plane-layered media. The first phase is constituted by metal; the second phase is constituted by semiconductor. By way of example, the following parameter values are selected: $\kappa_1$=99.314, $\sigma_1$=5·10$^6$, $\kappa_2$=0.637, $\sigma_2$=3.207·10$^4$. The value of ($\sigma_{xx}/\kappa_{xx}$) is rated for this ratio in the absence of thermoelectric phenomena – ($\sigma_{xx}/\kappa_{xx}$)/($\sigma_{xx}/\kappa_{xx}$)$_{lin}$.*

Exclusion from (17) $f_\lambda$ and $f_\mu$ functions leads to the general relationship interrelating values of $\sigma_e$, $\alpha_e$, $\kappa_e$ and independent on the specific medium structure [11]:

$$\alpha_e^2+\left(\frac{\kappa_2/\sigma_2-\kappa_1/\sigma_1}{\left(\alpha_1-\alpha_2\right)^2 T}+\alpha_1+\alpha_2\right)\alpha_e-\frac{\sigma_1\alpha_1\dfrac{\kappa_2}{T}\left(1+Z_2T\right)-\sigma_2\alpha_2\dfrac{\kappa_1}{T}\left(1+Z_1T\right)}{\sigma_1\sigma_2\left(\alpha_1-\alpha_2\right)}+\frac{1}{T}\frac{\kappa_e}{\sigma_e}=0\,. \tag{21}$$

Expression (21) can be considered a quadratic equation relative to $\alpha_e$. The limitation to $\sigma_e/\kappa_e$ relation follows from the requirement of positive discriminant of this equation. For the Wiedemann-Franz media ($\sigma_1/\kappa_1=\sigma_2/\kappa_2$) this limitation takes the following form:

$$\frac{\sigma_e}{\kappa_e}\geq\frac{\sigma}{\kappa}\frac{1}{1+\dfrac{1}{4}\dfrac{\sigma}{\kappa}\left(\alpha_1-\alpha_2\right)^2 T}\,. \tag{22}$$

Relationship (22) holds for two-phase inhomogeneous media of any structure, isotropic on the average. Comparison of (22) and (15) yields that self-dual media fulfill the least possible $\sigma_e / \kappa_e$ relation.

## Conclusions

The relation of effective electrical to thermal conductivity in the macroscopically inhomogeneous materials can considerably decrease with an increase in thermoelectric inhomogeneity. In turn, this decrease results in the decrease of effective figure of merit of inhomogeneous material. In general, the decrease of effective figure of merit in inhomogeneous media is substantially dependent on so-called intrinsic figure of merit.

This is the year of 100-th anniversary of outstanding physicist theoretician A.G. Samoylovich, theacher of the first of co-authors. A.G. Samoylovich achieved great success in studying the methods of energy conversion (see, for instance [1]). The same issues are dealt with in our publication devoted to the memory of A.G. Somoilovich.